\documentclass[12pt]{article} 
\begin{document}
\newcommand{\Eq}[1]{Eq.~(\ref{#1})}
\def\be{\begin{equation}}
\def\ee{\end{equation}} \def\bea{\begin{eqnarray}}
\def\eea{\end{eqnarray}}
\def\a{\alpha}
\def\b{\beta}
\def\e{\lambda}
\def\d{\delta}
\def\p{\psi}
\def\t{\nu}
\def\x{\xi}
\def\s{\sigma}
\def\G{\Gamma}
\def\o{\omega}

\title{Duality relation for frustrated spin models}
\author{D.-H.
Lee, Department of Physics \\ University of California, Berkeley,
California 94720 \\ and \\ F. Y. Wu, Department of Physics \\
Northeastern University, Boston, Massachusetts 02115}
\date{}
\maketitle

\begin{abstract}

We consider discrete spin models on arbitrary planar graphs and lattices
with frustrated interactions. We first analyze the Ising model
with frustrated plaquettes.
We use an algebraic approach to derive the result that
 an Ising
model with some of its plaquettes
frustrated has a dual which is an Ising model with an external field
$i\pi/2$ applied to the dual sites centered at frustrated
plaquettes.
In the case that all plaquettes are frustrated, this leads to the
known result that the dual model has a uniform
field $i\pi/2$ whose
partition function can be evaluated in the thermodynamic limit
for regular lattices.
  The analysis is extended to a Potts
spin glass with analogous results obtained.

\end{abstract}

\newpage

\section{The frustrated Ising model}

A central problem in the study of lattice-statistical problems
is the consideration of
frustrated spin systems (see, for example, \cite{v} - \cite{suzuki}).
  A particularly useful tool in the study of spin systems is the
consideration of duality relations (see, for example, \cite{wuwang,savit}).
Here we apply the duality consideration to frustrated discrete
 spin systems.

We consider first the Ising model on an {\it arbitrary}
planar graph $G$. A
 planar graph  is a collection of vertices and (noncrossing) edges.
   Place Ising spins at vertices of $G$ which interact
 with competing interactions along the edges.  Denote the
interaction between sites $i$ and $j$ by $-J_{ij}= - S_{ij} J$
where $S_{ij} =\pm 1$ and $J>0$. Then the Hamiltonian is
\be {\cal
H}(\s;S) =-\sum_{<i,j>}{ S_{ij}J\ \s_i\s_j}  \label{hamiltonian}
\ee
where $\s_i=\pm 1$ is the spin at the site $i$ and the
summation is taken over all interacting pairs.

The Hamiltonian
(\ref{hamiltonian}) plays an important role in condensed matter
physics and related topics. Regarding $S_{ij}$ as a quenched
random variable governed by a probability distribution, the
Hamiltonian (\ref{hamiltonian}) leads to the Edwards-Anderson
model of a spin glasses \cite{EA}. By taking a different $S_{ij}$,
the Hamiltonian becomes the Hopfield model of neural networks
\cite{hopfield}. Here, we consider  the Hamiltonian
(\ref{hamiltonian}) with fixed plaquette frustrations.

Let $G$
have  $N$ sites and $E$ edges.  Then it  has
\be
N^* = E +2 -N\
\hskip 1.5cm {\rm (Euler \>\> relation)} \label{euler}
\ee
faces,
including  one infinite face containing the infinite region and
$N^*-1$ internal faces which we refer to as plaquettes. The parity
of a face is the product of the edge $ S_{ij}$ factors around the
face which can be either $+1$ or $-1$. A face   is {\it
frustrated} if its parity  is $-1$. An Ising model is frustrated
if any of its plaquettes is frustrated, and is fully (totally)
frustrated if every plaquette is frustrated.
The fully frustrated model is also known as the odd model
of the spin glass \cite{v}. As the parity of the
infinite face is the product of the parities of all plaquettes,
the parity of the infinite face  in a totally frustrated Ising
model is  $-1$ for $N^*=$ even and $+1$ for $N^*=$ odd. An example
of a full frustration is the triangular model with $S_{ij} = -1$
for all nearest neighbor sites $i,j$.

The values of parity associated with all plaquettes defines a
``parity configuration'' which we denote by $\G$. The set of
interactions $\{S_{ij}\}$ corresponding to a given $\G$  is not
unique. For the triangular model, for example, any $\{S_{ij}\}$
which has either one or three $S_{ij} = -1$ edges around every
plaquette is totally frustrated. For a given $\{S_{ij}\}$   and
$\G$, the partition function is the summation
\be Z\big(
\{S_{ij}\}\big) = \sum_{\s_1=\pm 1} \cdots \sum_{\s_N=\pm 1} \prod
_{E}e^{ S_{ij}J\ \s_i\s_j}\ \label{spinglass} \ee
where the
product is taken over the $E$
edges of $G$.

\subsection{Gauge transformation}

A {\it gauge} transformation is site-dependent redefinition of the
up/down spin directions. Mathematically a gauge transformation
transforms the spin variables according to  \cite{fhs}
\be
\s_i \to
\s_i'=w_i\ \s_i,~~~i=1,..,N.
\ee
In the above if $w_i=+1$ the
original definition of up/down spin directions is maintained, and
if $w_i=-1$ the definitions of up/down are exchanged. Under the
gauge transformation the $S_{ij}$ in \Eq{hamiltonian} transform as
follow
\be
S_{ij} \to S_{ij}' = w_i\ S_{ij}\ w_j \hskip
1cm~~\forall~i,j.\label{gauge}
\ee
Since $w_i^2=1$, we have \be
{\cal H} (\s; S) = {\cal H} (\s'; S'). \ee Clearly, the gauge
transformation (\ref{gauge}) leaves the parity configuration $\G$
unchanged, i.e., \be \prod_{{\rm face}}S_{ij}=\prod_{{\rm
face}}S'_{ij}\hskip 1cm~~\forall~{\rm face}.\label{invg}\ee

For each parity configuration $\G$ there are $2^{N-1}$ different
$\{S_{ij}\}$ patterns consistent with it. To see that we note in
\Eq{gauge} each of the $2^N$ choices of $\{w_i\}$ leads to a new
$\{S_{ij}'\}$ except the negation of all $w_i$ which leaves
$\{S_{ij}\}$ unchanged. Conversely, any two sets of interactions
$\{S_{ij}\}$  and $\{S_{ij}'\}$ for the same $\G$ are related by a
gauge transformation which can be constructed as follows. Starting
from any spin, say, spin 1, assign the value $w_1=   +1$. One next
builds up the graph  by adding one site (and one edge) at a time.
To the site $2$ connected to $1$ by the edge $\{12\}$, one assigns
the factor $w_2 = w_1\ S_{12}\ S'_{12}$ which yields $w_1\ S_{12}\
w_2 = S_{12}' $ consistent to (\ref{gauge}). Proceeding in this
way around a plaquette until  an edge, say $\{n1\}$, completes a
plaquette.  At this point one has \be w_n\ S_{n1}\ w_1
=\Bigg(\prod_{\rm plaquette} S_{ij} \Bigg) \Bigg(\prod_{\rm
plaquette} S_{ij}' \Bigg) S_{n1}' =S_{n1}' \ee which is again
consistent to (\ref{gauge}). Continuing in this way one constructs
all $w_i$ which transform $\{S_{ij}\}$  into $\{S_{ij}'\}$.   Note
that if we had started with $w_1=-1$, we would have resulted in
the negation of all $w_i$. Thus, the bijection between the
$2^{N-1}$ sets $\{S_{ij}\}$  and $2^{N-1}$ gauge transformations
is one-one.

In addition to \Eq{invg}, the gauge transformation also leaves the
partition function invariant \cite{nish1,nish2}, i.e.
\bea
Z\Big(\{S_{ij}\}\Big)=\sum_{\{\s\}}\prod_E e^{S_{ij}J\s_i\s_j} =
\sum_{\{\s'\}}\prod_E
e^{S'_{ij}J\s'_i\s'_j}=Z\Big(\{S'_{ij}\}\Big).\label{invpart}
\eea
As a result the partition function only depends on $\G$ and we can
rewrite ({\ref{spinglass}) as \be Z(\G) =2^{-(N-1)}
\sum_{\{S_{ij}\}}Z\Big(\{S_{ij}\} \Big) \label{spinglass1} \ee
where the summation is over all $2^{N-1}$ distinct $\{S_{ij}\}$
consistent with the parity configuration $\G$ for the same
partition function. This expression of the partition function is
used to derive the duality relation in ensuing sections.

\subsection{The fully frustrated Ising model}
For our purposes
it is instructive to consider first the case of full frustration.
Duality properties of fully frustrated model have previously been considered
by a number of authors \cite{fhs,suzuki} for regular lattices.
 We present here an alternate formulation
applicable to   arbitrary
graphs and arbitrary frustration.

The graph $D$ dual to $G$ has $N^*$ sites each residing in a face
of $G$, and $E$ edges each intersecting an edge of $G$.
We restrict to $N^*=$ even so that all faces of $G$ including the
infinite face are frustrated. This restriction has no effect on
the taking of the thermodynamic limit in the case of regular lattices.
Since the
signs $S_{ij}$ around each face are subject to the constraint
$\prod S_{ij} = -1$, we introduce in the summand of
(\ref{spinglass1}) a face factor $(1-\prod S_{ij})/2$ and sum over
$S_{ij}=\pm 1$ independently. Similarly, writing
$\s_{ij}=\s_i\s_j$, we can replace  summations over $\s_i= \pm 1$
in (\ref{spinglass1}) by $\s_{ij} = \pm 1$ by introducing a factor
$(1+\prod \s_{ij})/2$ to each face. Thus, (\ref{spinglass1})
becomes \be Z_{FF}= \Bigg[\frac{2\cdot 2^{-(N-1)}}{2^{2N^*}}
\Bigg] {\sum_{\{\s_{ij}\}}} \ {\sum_{\{S_{ij}\}}} \Big[\prod _E
e^{S_{ij} \s_{ij} J} \Big] \prod_{\rm face} \Big[(1+\prod \s_{ij})
(1-\prod S_{ij})\Big] \label{part1} \ee where the subscript FF
denotes full frustration, and the  extra factor 2 in (\ref{part1})
is due to the $2\to 1$ mapping from $\s_i$ to $\s_{ij}$.

For a face having $n$ sides we rewrite the face factors as
\begin{eqnarray}
1+\prod \s_{ij} &=& \sum _{\mu=\pm } \prod
F(\s_{ij};\mu) \label{constraint1} \\ 1-\prod S_{ij} &=&  \sum
_{\t=\pm } \prod G(S_{ij};\t) \label{constraint2}
\end{eqnarray}
where each product has  $n$ factors \bea F(\s ;\mu) &=& \d_{\mu
+}+ \s\ \d_{\mu -} \nonumber\\ G(S ;\t) &=&  \d_{\t+}+ S\ \o_n \
\d_{\t -}\ , \label{fg} \eea $\d$ is the Kronecker delta function
and $\o_n = (-1)^{-1/n} = e^{-i\pi/n}$.

We now regard  $\mu$ and $\t$ as indices of two Ising spins
residing at each dual lattice site. After carrying out summations
over $\s_{ij}$ and $S_{ij}$, the partition function (\ref{part1})
becomes \be Z_{FF}  = 2^{-E-N^*}\sum_{\{\mu\}} \sum_{\{\t\}}
\prod_{E} B(\mu,\t; \mu', \t') \label{dualpart} \ee where we have
made use of the Euler relation (\ref{euler}) and $B$ is a
Boltzmann factor  given by
\be
B(\mu,\t; \mu ',\t') = \sum_{\s
=\pm 1}\sum_{S =\pm 1}e^{S\s J} F(\s;\mu) F(\s; \mu') G(S;\t)
G(S;\t').\label{dualB}
\ee
Here $G(S;\t')$  is given  by
(\ref{fg}) with $\o_n \to \o_{n'} = e^{-i\pi/n'}$ and the two
faces containing spins $\{\mu,\t\}$ and $\{\mu',\t'\}$ have
respectively $n$ and $n'$ sides.

Substituting (\ref{fg}) into (\ref{dualB}) and making  use of the
identities
\bea \d_{\mu+}\d_{\mu'+} +\d_{\mu-}\d_{\mu'-} &=&
(1+\mu \mu')/2\nonumber \\ \d_{\mu+}\d_{\mu'-}
+\d_{\mu-}\d_{\mu'+} &=& (1-\mu \mu')/2,
\eea
one obtains
\be
B(\mu,\t; \mu ',\t') =2A (1+\mu \mu') \cosh J + 2B (1-\mu\mu')
\sinh J \label{dualB1}
\ee
where
\bea A &=& \d_{\t+}\d_{\t'+}
+\o_n \o_{n'} \d_{\t-}\d_{\t'-} \nonumber \\ B &=& \o_n \
\d_{\t-}\d_{\t'+} + \o_{n'} \d_{\t+}\d_{\t'-}
\eea
Number the four
states $\{\mu,\nu\}= \{ +,+\},\{ -,-\},\{ -,+\},\{ +,-\}$
respectively by $1,2,3,4$.   The Boltzmann factor (\ref{dualB1})
can be conveniently written as a $4\times 4$ matrix \be B(\mu,\t;
\mu ',\t')= \pmatrix{ B_{11} & B_{12}&0&0\cr B_{21}& B_{22} & 0 &
0\cr 0&0&B_{11} & B_{12} \cr 0&0& B_{21}& B_{22}} \ee where \bea
B_{11} &=& 4\cosh J, \hskip 2cm B_{12} = 4\ \o_{n'}\ \sinh J,
\nonumber \\ B_{21} &=&4\ \o_n\ \sinh J, \hskip 1.3cm B_{22} = 4\
\o_n\ \o_{n'}\ \cosh J\ . \eea Thus the partition function of the
$\{\mu,\t\}$ spin model is twice that of an Ising model on the
dual lattice. The exchange coupling constant $K$ and the magnetic
field $h$ in the dual model are determined by
\bea
B_{11} &=& D
e^{K+(h/n)+(h'/n')}, \hskip 1.2cm B_{12} = D e^{-K+(h/n)-(h'/n')}
\nonumber \\ B_{21} &=& D e^{-K-(h/n)+(h'/n')}, \hskip 1cm B_{22}
= D e^{K-(h/n)-(h'/n')} \ .
\eea
Here $n$ and $n'$ are the number
of edges incident at the two dual sites respectively.

The solution of the above equations gives
\bea
e^{-2K} &=&\tanh J
>\ 0, \hskip
0.8cm D= 4(\o_n\ \o_{n'})^{1/2}\sqrt {\sinh J\cosh J} \nonumber\\
e^{2(h/n)} &=& 1/\o_n=e^{i\pi/n}, \hskip 0.8cm e^{2(h'/n')} =
1/\o_{n'}=e^{i\pi/n'}\ , \label{transf}
\eea
or equivalently
\be
K=-\frac{1}{2}\ln\Big(\tanh{J}\Big)~~{\rm
and}~~h=h'=\frac{i\pi}{2}.\ee
Thus, we have established the
equivalence \be Z_{\rm FF} (J) = 2^{N-1}\ i^{N^*} (\sinh J\cosh
J)^{E/2}\ Z^{(D)}_{\rm Ising} \Big(i{{\pi}\over 2},
K\Big)\label{dualfield}
\ee
where $Z^{(D)}_{\rm Ising}
\Big(i{{\pi}/ 2}, K\Big)$ is the partition function of a
ferromagnetic Ising model on $D$ with interactions $K>0$ and an
external field $i\pi/2$. In writing down (\ref{dualfield}) we have
made use of the identity $2\cdot 2^{-(E+N^*)}4^E = 2^{N-1}$ and
the fact that $(\o_n\ \o_{n'})^{E/2} = (-i)^{N^*}=i^{N^*}$ for
$N^*=$ even.

\noindent
Remarks:

1. The duality relation (\ref{dualfield})
has previously been obtained
by Fradkin {\it et al.} \cite{fhs}, and for the
square lattice by Suzuki \cite{suzuki} and S\"ut\~o \cite{suto},
and by Au-Yang and Perk \cite{auyangperk} in another context.

2. The duality relation (\ref{dualfield}) is different
from the Kadanoff-Ceva-Merlini scheme  \cite{kc,m} of  replacing $K$ by $K +i\pi/2$
(corresponding to $J<0$ in (\ref{transf}))  in the ferromagnetic
model.  Suzuki \cite{suzuki}
has made the explicit use of the Kadanoff-Ceva-Merlini scheme
in deriving  (\ref{dualfield}) for the square lattice.
 For  fully frustrated systems the Suzuki method can be extended to
 any graph whose dual admits dimer coverings.

3. The duality relation (\ref{dualfield}) holds for a
fixed $\{S_{ij}\}$ without probability considerations, and
therefore differs intrinsically from that of a spin glass obtained
recently by Nishimori and Nemoto \cite{nn} using a replica
formulation.

4. The duality relation (\ref{dualfield}) which holds for any lattice
appears to support the suggestion \cite{forgacs} that all fully frustrated
Ising models belong to the same universality class.

\subsection{The thermodynamic limit} The partition function
(\ref{dualfield}) for an Ising model in a uniform field $i\pi/2$
can be exactly evaluated for regular lattices. Defining the per-site ''free energy"
 \be
 f =  \lim_{N^* \to\infty} \frac{1}{N^*}\ln Z^{(D)}_{\rm Ising}
\Big(i{{\pi}\over 2}, K\Big),
\ee
 Lee and Yang \cite{leeyang}
have obtained a closed form expression of $f(K)$ for the square
lattice. Their result, which was later derived rigorously by McCoy
and Wu \cite{mccoywu} and others \cite{m,linwu}, is
\be
f=i{\pi\over 2}+C +{1\over
16\pi^2}\int_{-\pi}^{\pi}d\theta\int_{-\pi}^{\pi}d\phi
\ln[z+z^{-1}+2-4\cos\theta\cos\phi], \label{pffield}
\ee
where $C=
[\ln(\sinh 2K)]/2,\ z=e^{-4K}$. The free energy (\ref{pffield}),
which is the same as that obtained by Villian \cite{v}, is
analytic at all nonzero temperatures.

The solution for the triangular Ising model in a field $i\pi/2$ has also
been deduced previously  \cite{linwu,luwu}. However, it can also
be obtained most simply by observing that the honeycomb lattice,
which is the dual of the triangular lattice, has a coordination
number 3. It follows that we can recast the field Boltzmann
weights as $e^{i\pi\s_j/2} = i \s_j=i \s_j^3$ and redistribute the
$\s_j^3$ factor at site $j$ to its three incident  edges. Then,
as pointed out by Suzuki \cite{suzuki},
each edge can be associated with a factor $i\s_i\s_je^K =
e^{(K+i\pi/2)\s_i\s_j}$ and the desired solution can be obtained
from that of the {\it zero-field} honeycomb lattice with the
simple replacement $K\to K+i\pi/2$. This gives
\be
f=i{\pi\over
2}+C+{1\over 16\pi^2}\int_{-\pi}^{\pi}d\theta\int_{-\pi}^{\pi}d\phi
\ln\Big[(1+e^{4K})^2 +4\cos\phi(\cos\theta+\cos\phi) \Big],
 \ee
where $C=[\ln( 2\sinh 2K)]/2$ and $K$ is the Ising interaction on the
honeycomb lattice. Again, there is no finite
temperature phase transition.

\subsection{Arbitrary plaquette
parities}

In a similar fashion one can extend the above analysis to Ising
models with arbitrary face parities.  All steps of previous
subsections can be carried through except that for faces which are
not frustrated we must replace $\o_n$ by $ 1$ at the corresponding
dual sites. This results in a zero field (instead of a field
$i\pi/2$) at these sites. Thus, for an Ising model with arbitrary
parity configuration $\G$, its dual model has fields $0$ and
$i\pi/2$, respectively, at sites  in faces of parity $+1$ and
$-1$. Explicitly, we find
\be
 Z(\G) = 2^{N-1} (-i)^{N_F} (\sinh J
\cosh J)^{E/2} Z^{(D)}_{{\rm Ising}}\Big(\{h_j\},K\Big)
\label{arbfrus}
\ee
where the dual partition function is
 \be
Z^{(D)}_{{\rm Ising}}\Big(\{h_j\},K\Big)=\sum_{\{\mu_i\}} \prod_E
e^{K\mu_i\mu_j}\prod_{{\rm face}} e^{h_i\mu_i}\label{arbfrus1}\ .
\ee
 Here, $N_F$ is the number of frustrated faces and the
external field at site $j$ is $h_j=i\pi/2$ or $0$ depending on
whether the face associated with the site is frustrated or not.

\noindent
Remarks.

1. The duality relation (\ref{arbfrus1}) for Ising models with
arbitrary frustrated plaquettes can be found as contained implicitly
in \cite{fhs}.

2. By writing $e^{i\pi\s/2} = i\s$ in the dual partition
function, we see that the partition function of an Ising model
with $p$ frustrated faces is dual to a $p$-spin Ising correlation
function in zero field. Particular, the $p=2$ correlation problem
has been studied in details \cite{mccoywu} which now leads to a
wealth of information on the correlation of two frustrated
plaquettes.

\section{Potts spin glass}
Our analysis can be extended to a $q$-state spin model, the Potts
spin glass. First, we recall the definition of a chiral Potts
model. The chiral Potts model, which was first considered in
\cite{wuwang}, is a discrete $q$-state spin model with a cyclic
Boltzmann factor $\Lambda (\x,\ \x')=\Lambda (\x-\x')$ between two
spins at sites $i$ and $j$ and in states $\x_i$ and $ \x_j = 0,
1,\cdots,q-1$. The interactions are $q$-periodic, namely, the
Boltzmann factor satisfies \be U(\x+q) = U(\x). \ee The
interaction can be symmetric, namely, $U(\x)=U(-\x)$,  as in the
case of the standard Potts model \cite{potts}, but in our
considerations this  needs not  be the case.

A Potts spin glass is a chiral Potts model with  random
interactions. To describe the randomness one  introduces edge
variables $\e_{ij} =\e_{ji} = 0, 1, \cdots, q-1$ and considers the
partition function \cite{nn,ns,jp}
\be
Z_{\rm
Potts}\Big(\{\e_{ij}\}\Big) =\sum _{\x_i=0}^{q-1}  \prod_E U(\x_i
- \x_j +\e_{ij})\ . \label{PottsZ}
\ee
Note that if $U$ is
symmetric and $q=2$, the partition function (\ref{PottsZ}) reduces
to (\ref{spinglass}). A plaquette has ``flux'' $r$
($=0,1,2,\cdots, q-1$) if \cite{ltg}
\be\sum_{\rm plaq} \e_{ij}=r~~({\rm
mod~q}).\ee  A set of
$\{\e_{ij}\}$ leads to a flux configuration $\G$, which is
specified by the values of the flux for all faces.

\subsection{Gauge transformation}
A gauge transformation for the Potts spin glass is the mapping
\bea
\x_i &\to& \x_i'=\x_i+\theta_i \nonumber \\ \e_{ij} &\to&
\e_{ij}' =\e_{ij} +\theta_i-\theta_j,
\eea
where $\theta_i =
0,1,\cdots,q-1$. It is clear that this mapping leaves the flux
configuration $\G$ unchanged, i.e.,
\be
\sum_{{\rm plaq}}\e_{ij}
=\sum_{{\rm plaq}}\e'_{ij}.
\ee
Since a global change of all $\x_i$
by the same amount preserves $\{\e_{ij}\}$, the total number of
distinct $\{\e_{ij}\}$ consistent with a particular flux
configuration $\G$ is $q^{N-1}$. Conversely, any two sets of
$\{\e_{ij}\}$  and $\{\e_{ij}'\}$ giving rise to the same flux
configuration are related through a gauge transformation. To see
this we start from a arbitrarily chosen site, say site 1, and set
$\theta_1=0$. Next we assign $\theta_2 = \theta_1 + \e_{12} -
\e_{12}'$ to site 2 connected to site 1 by an edge. Continuing in
this way as in the Ising case, one eventually determines a set of
$\theta_i$ which transforms $\{\e_{ij}\}$ into $\{\e_{ij}'\}$ ,
and vice versa. The bijection between the $q^{N-1}$ configurations
of $\e_{ij}$ and gauge transformations for a given $\G$ is
one-one.

In addition to leaving the flux configuration unchanged, gauge
transformation also leaves the partition function invariant,
namely, \be Z_{\rm Potts}\Big(\{\e_{ij}\}\Big)=Z_{\rm
Potts}\Big(\{\e'_{ij}\}\Big).\ee Thus, in analogous to
(\ref{spinglass1}), we have \be Z_{\rm Potts}(\G) = q^{-(N-1)}
\sum_{\{\e_{ij}\}} Z\Big( \{\e_{ij}\}\Big)\label{Pottsglass1} \ee
which is used to derive a duality relation. Again, the sum in
\Eq{Pottsglass1} run through all $\{\e_{ij}\}$ consistent with a
given flux configuration $\G$.

\subsection{Duality relation}
In the Potts partition function (\ref{PottsZ}), write $\x_{ij}
=\x_i-\x_j$ and to each face having a flux $r$ introduce two
factors
\bea
{1\over q} \sum_{\mu =0}^{q-1} e^{-i2\pi \mu
(\x_{12}+\x_{23}+\cdots +\x_{n1})/q} &=& 1, \quad {\rm
if\>\>}\x_{12} +\cdots+\x_{n1} = 0 \>\>({\rm mod}\>q) \nonumber \\
&=& 0, \quad {\rm otherwise} \nonumber \\ {1\over q} \sum_{\t
=0}^{q-1} e^{-i2\pi \t(\e_{12}+\e_{23}+\cdots +\e_{n1}-r)/q} &=&
1, \quad {\rm if\>\>}\e_{12} +\cdots +\e_{n1} =r  \>\>({\rm
mod}\>q) \nonumber \\ &=& 0, \quad {\rm otherwise}\ .
\label{constraint2}
\eea
This permits us to sum over $\x_{ij}$ and
$\e_{ij}$ independently. Thus, in analogous to (\ref{dualpart}),
we obtain
\be
Z_{\rm Potts}(\G) = q^{-E-N^*}
\sum_{\{\mu\}}\sum_{\{\nu\}} B(\mu,\nu;\mu',\nu')
\label{dualpartP} \ee where \bea &&B(\mu,\nu;\mu',\nu') =
\sum_\lambda \sum_\xi U(\xi + \lambda) \nonumber \\ &&\hskip 1cm
\times \ {\rm exp} \Bigg[ \frac{i2\pi}q \Bigg( -(\mu-\mu')\xi
-(\nu-\nu')\lambda +{{r\nu}\over { n}}+{{r'\nu'}\over {n'}}\Bigg)
\Bigg]  \label{BP}
\eea
and $n,n'$ are the numbers of sides of the
two plaquettes containing $\{\mu,\nu\}$ and $\{\mu',\nu'\}$,
and flux $r$ and $r'$
respectively.

Carry out the summations in (\ref{BP}) after introducing the
Fourier transform
\be
U(\x+\e) = {1\over q} \sum_{\eta=0}^{q-1}\
\Lambda(\eta)\ e^{i2\pi\eta(\x+\e)/q} \label{eigenvalue}
\ee
where
$\Lambda(\eta)$ are the eigenvalues of the matrix $U$
\cite{wuwang}. One obtains \be B(\mu,\nu;\mu',\nu') =q \
\delta_{\mu-\mu', \nu-\nu'}\ \Lambda(\nu-\nu') \  e^{i2 \pi
r\nu/qn}\ e^{i 2\pi r'\nu'/qn'}\ . \label{BP1} \ee In the above
$\delta_{\mu-\mu', \nu-\nu'}$ sets $\mu-\mu'$ to $\nu-\nu'$ (mod
q).

The substitution of (\ref{BP1}) into (\ref{dualpartP}) followed by
summing over $\mu$  now yields the result
 \be
Z_{\rm Potts}(\G) =
q^{1-N^*}Z_{\rm Potts}{^{(D)}}(\{h_j\},\Lambda)\label{dualP} \ee
 where
\be
Z_{\rm Potts}{^{(D)}} (\{h_j\},\Lambda)= \sum_{\{\nu_i\}}\Big[ \prod_E
\Lambda(\nu_i, \ \nu_j) \Big]\Big[\prod_{{\rm face}} e^{h_j
\nu_j }\Big]\ . \label{dualfieldP}
\ee
is the partition
function of a chiral Potts model on the dual graph $D$, which
generalizes (\ref{arbfrus}) to Potts spin glasses.  The dual
chiral Potts model  has Boltzmann weights $\Lambda(\mu_i,\ \mu_j)=
\Lambda(\mu_i-\mu_j)$ and external fields
\be
h_j=i\ {{ 2\pi r_j}\over q}\ , \hskip .6cm j=1,2,\cdots,N^*, \quad
r_j = 0,1,\cdots,q-1 \ee
 on the
spin in plaquette $j$ which has a flux $r_j$.
 When $r_j=0$ for all $j$,  (\ref{dualP}) reduces
to the duality relation for the zero-field chiral Potts model given
by Eq. (13) in \cite{wuwang}.

\section{Summary
and acknowledgments}

We have obtained  duality relations for planar Ising and chiral
Potts models on arbitrary graphs
and with fixed plaquette parity or flux configurations.
Our main results are the equivalences (\ref{dualfield}) for the
fully frustrated Ising model, (\ref{arbfrus}) for the Ising model
with arbitrary plaquette parity, and (\ref{dualP}) for the chiral
Potts model with arbitrary flux configurations.  In all cases the
dual models have pure imaginary fields applied to spins in
plaquettes that are frustrated and/or having a nonzero flux.

We thank  H. Kunz for a conversation which motivated this work.
We are also indebted to C. Henley, H. Nishimori, J. H. H. Perk,
L. Pryadko, and A. S\"ut\~o  for calling our attention to relevant
references.

Work has been supported in part by NSF Grants DMR-9971507 and
DMR-9980440.

\end{document}